\documentclass[preprint]{aastex61}
\usepackage{amsmath,booktabs,graphicx,subfigure,amssymb,latexsym,graphics,epsfig,exscale,url}

\begin{document}

\title{ALMA's Polarized View of 10 Protostars in the Perseus Molecular Cloud}

\correspondingauthor{Erin G. Cox}
\email{egcox2@illinois.edu}

\author{Erin G. Cox}
\affil{Department of Astronomy, University of Illinois at Urbana-Champaign,
1002 W. Green St.,
Urbana, IL 61801, USA}

\author{Robert J. Harris}
\affiliation{National Center for Supercomputing Applications,
University of Illinois at Urbana-Champaign, 1205 W Clark St, Urbana, IL 61801} 

\affil{Department of Astronomy, University of Illinois at Urbana-Champaign,
1002 W. Green St.,
Urbana, IL 61801, USA}

\author{Leslie W. Looney}
\affil{Department of Astronomy, University of Illinois at Urbana-Champaign,
1002 W. Green St.,
Urbana, IL 61801, USA}

\author{Zhi-Yun Li}
\affiliation{Department of Astronomy, University of Virginia, 530 McCormick Rd, Charlottesville, VA 22903, USA}
\author{Haifeng Yang}
\affiliation{Department of Astronomy, University of Virginia, 530 McCormick Rd, Charlottesville, VA 22903, USA}

\author{John J. Tobin}
\affiliation{Homer L. Dodge Department of Physics and Astronomy,  Oklahoma University, 440 W. Brooks St.
Norman, OK 73019, USA}
\affiliation{Leiden Observatory, Leiden University, P.O. Box 9513, 2000-RA Leiden, The Netherlands}

\author{Ian Stephens}
\affiliation{Harvard-Smithsonian Center for Astrophysics, 60 Garden St, Cambridge, MA 02138, USA}

\begin{abstract}
We present 870 $\mu$m ALMA dust polarization observations of 10 young Class 0/I protostars in the
Perseus Molecular Cloud. At $\sim0.35\arcsec$ (80 au) resolution, all of our sources show some degree of polarization, with most (9/10) 
showing significantly extended emission in the polarized continuum. Each source has incredibly intricate polarization
signatures. In particular, all three disk-candidates have polarization vectors roughly along the minor axis, 
which is indicative of polarization produced by dust scattering. 
On $\sim$ 100 au scales, the polarization is at a relatively low level ($\lesssim 1\%$) and is quite ordered. In sources
with significant envelope emission, the envelope is typically polarized at a much higher ($\gtrsim 5\%$) level and has a far more disordered morphology. 
We compute the cumulative probability distributions for both the small (disk-scale) and large (envelope-scale) polarization percentage. We find that the two are intrinsically different, even after accounting for the different detection thresholds in the high/low surface 
brightness regions. We perform Kolmogorov-Smirnov and Anderson-Darling tests on the distributions of angle offsets of the polarization from the
outflow axis. We find disk-candidate sources are different from the non-disk-candidate sources. We conclude that the polarization on the 100 au scale is consistent with the signature of dust scattering for disk-candidates and that the polarization on the envelope-scale in all sources
may come from another mechanism, most likely magnetically aligned grains. 
\end{abstract}

\section{Introduction} \label{sec:intro}
High-resolution, sub-/millimeter dust continuum observations of protostars preferentially probe grains
that are $\sim 0.1-\mathrm{few}$ mm-- the pebble-sized rocks that are 
the first step towards planetary bodies. 
These grains are responsible for continuum polarization in protostars.
A few mechanisms could be responsible; each reveals a different facet of the
protostar's nascent environment. These mechanisms  can depend on the grain population, opacity and, through these, on the size-scale observed 
\citep[e.g., envelope/disk,][]{la07,ka15,ka16,ta17}. Previous polarization surveys of protostars 
\citep{ch13,hu14,zh14} have been at relatively low resolution and only probed these objects' 
nascent envelope. High resolution observations of the environment close 
to the protostar \citep[$\sim$ few hundred au, such as those done by][in Serpens]{hu17}
are necessary to disentangle the emission from the envelope and disk and therefore
understand how grains change in the envelope-disk transition.

Traditionally, dust continuum polarization observations are used as a proxy for the magnetic field. 
Dust grains align themselves perpendicularly to the local magnetic field and consequently emit polarized
thermal emission \citep{la07}. 
Observationally constraining the field's morphology and strength is 
important in star formation on all scales. In the $\sim$ 0.1pc core where protostars will form, 
magnetic fields, if strong enough, control the initial collapse 
through both magnetic pressure support and ambipolar diffusion \citep[e.g.,][]{sh87}, directly influencing
the timescale on which stars can form. Once the cloud has begun its collapse, infalling 
material can travel along these field lines from large (cloud) scales to the small (disk)
scales \citep{li14a}. On small scales, magnetic fields are thought to play an important role in the outflows,
 jets, wind launching, and disk accretion \citep[e.g.,][]{bl82,ba98}. In fact, in the youngest protostars 
(Class 0 sources; \citealt{an93}), these fields may create efficient magnetic braking, hindering large (R $\sim$ 100 au) 
disk growth until later evolutionary phases, such as the Class I phase.

Indeed, ideal MHD simulations have shown that magnetic braking can be strong in Class 0 sources \citep[e.g.,][]{me08,he08}.
These simulations show that when magnetic fields are
aligned with the infalling envelope's rotation axis, the magnetic braking is efficient enough to strip 
infalling material of angular momentum \citep{me08,he08}.
The angular momentum loss can hinder large disk formation in these sources \citep{li11}
until they are older Class I/II objects (e.g., \citealt{me09,an09,an10,da10}).  
Conversely, if the field is perpendicular or misaligned, magnetic braking is not as efficient, 
and a large disk may grow (e.g., \citealt{jo12}). Observations of the inferred magnetic field of 
Class 0 sources have shown this dependence of disk formation on field alignment to hold down to $\sim$ 1000-few hundred au \citep[e.g.,][]{hu14,se15}. 
Early analytic estimates of the spin-down time for rotating starless
cores embedded in a non-rotating, more tenuous medium might seem to contradict this result because they indicate more efficient 
magnetic braking in an aligned rotator instead of an orthogonal rotator \citep[e.g.,][]{mo85}. However, these analytic estimates did not
account for the change in the magnetic field configuration during the protostellar phase. Most of the protostellar phase was also missed in 
the pioneering 3D ideal MHD simulations of magnetic braking of \citet{ma04} and in non-ideal MHD 
simulations of \citet{ma16}, who did not find a significant effect of magnetic alignment on braking efficiency.

Testing these magnetic braking scenarios is complicated by the fact that other mechanisms unrelated to the magnetic field, such as self-scattering of the dust's thermal emission, can generate polarized continuum emission.
Self-scattering is inefficient on large cloud scales because the grains are 
too small ($a\sim\mu$m) to scatter millimeter-wavelength light efficiently. On the disk scale, where the grains
could be much larger, new theoretical studies (e.g., \citealt{ka15,ka16,ya16a,ya16b}) have shown that 
scattering may in fact produce the polarization observed in some protostellar disks \citep{ra14,st14,se15,co15,fe16,ka16b}. 
This mechanism may be useful because it can be used to probe the size of the dust 
grains observed, independent of the spectral index ($\alpha$), using the scattering cross-section as a proxy \citep{ka15,ka16}. 
This may be a good alternative to measuring the grain size
using the spectral index because $\alpha$ is degenerate in areas of high optical depth or low temperatures, such as young disks. 
Since grains are expected to grow more easily in a dense,
rotationally supported disk than in a rapidly infalling envelope, grains are more likely to exhibit scattering-induced polarization
in disks than in envelopes. Polarization in disks and envelopes may therefore have different origins.

Another mechanism for producing polarized emission is  direct alignment of the grains via radiation pressure. \citet{la07b} first discussed this mechanism, and \citet{ta17} has shown 
that, in a disk, instead of aligning with an external magnetic field, large grains may align with their short axes along the 
direction of the radiative flux anisotropy. This would produce an azimuthal polarization pattern in a disk.
The aligning photons may be in the far-infrared and may not come directly from the star. \citet{ka17} and \citet{st17b} 
have demonstrated that scattering and radiative alignment
dominate the emission at 850 $\mu$m and 3 mm, respectively, in the Class I/II source HL Tau.

While each of the three scenarios predicts different detailed structures in the polarization emission
in disks/envelopes, previous observations have not been able to disentangle them due to comparatively poor 
resolution ($\gtrsim 2.5\arcsec$, $\sim$ 400 au to 1000 au depending on the cloud), leading to disk/envelope confusion.
High-resolution observations are needed to accurately probe each structure.
In this article we present Atacama Large Millimeter/submillimeter Array (ALMA) 870 $\mu$m dust
continuum polarization observations of 10 Class 0/I protostars in the Perseus Molecular Cloud.

\section{Observation and Sample Selection} \label{sec:obs}

\subsection{Sample Selection}

Our target selection was constructed from our Very Large Array Nascent Disk and Multiplicity (VANDAM) survey \citep{to15,to16}. The VANDAM sources 
include all 94 identified protostellar systems in the Perseus Molecular Cloud \citep[$d \sim 230$ pc;][]{hi08}; as part of the survey, 
they were imaged at both Ka (8 mm/1 cm) and C (4/6 cm) bands.  Of these, we have preliminarily defined some of 
the sources as being ``disk-candidates,'' which means that they are resolved (perpendicular to the outflow direction when known) and 
that their emission profiles were fit by a model of a self-similar, 
viscously evolving density profile with a reasonable temperature profile \citep[see ][]{se16}; this excluded 
some extended sources \citep[see ][]{se17}.   For this paper, we started with the full VANDAM sample, then 
chose the brightest 25 at 220 GHz from the Submillimeter Array (SMA) Mass Assembly of Stellar Systems and their Evolution (MASSES) survey (private 
communication, Katherine Lee).   Of those, we selected the expected brightest in ALMA Band 7 assuming 
reasonable values for the 8 mm - 870 $\mu$m spectral index.  Finally, we only observed the 10 sources with no 
previous polarimetric observations in the millimeter/centimeter bands.   Of these 10 sources, only 3 protostars 
(Per-emb-11, Per-emb-14 and Per-emb-50) were identified as Class 0/I disk-candidates in \citet{se17}.   
The other 7 sources (hereafter, ``non-disk-candidates"), whose 8 mm emission could not be fit in \citet{se17}, included two 
(Per-emb 2 and 18) whose morphologies resemble a fragmenting disk, but could not be fit by the model, and one close binary (Per-emb 5).

\subsection{Observations and Data Reduction}
The data were taken using ALMA Band 7 on 2016, July 17 in configuration C40-4 (project code 2015.1.01503.S; P.I. Erin Cox). 
Our observations were taken in full polarization mode, with 4 spectral windows tuned for continuum observations.
The observations took a total of 2.6 hours, with $\sim$8 minutes spent per source. Baselines for
the C40-4 configuration ranged from 15 to 704 meters, corresponding to a maximum recoverable 
scale of $\sim 7.2 \arcsec$ and a resolution of $\sim 0.35\arcsec$. The 4 spectral windows were centered on 337.5, 339.4, 347.5, and 349.5 GHz. The
sources J0336+3218, J0238+1636, J0237+2848, and J0510+1800 were used to calibrate the phase, flux, bandpass, and 
polarization, respectively. J0319+3101 was used as the check calibrator for the phase transfer. 
The overall amplitude calibration uncertainty at Band 7 for ALMA is 10\%; we only report the statistical uncertainties 
on the flux densities reported in this paper. The absolute calibration uncertainty in the polarization position angle 
is $\sim 0.4^{\circ}$ \citep{na16}, which is smaller than the statistical uncertainties for these sources.

These observations were reduced manually by data analysts at the National Radio Astronomy Observatory (NRAO), using
the Common Astronomy Software Applications (CASA) package \citep{mc07}, version 4.7.0. 
These calibrations included first applying a priori calibrations, such as baseline corrections and 
phase corrections from water vapor radiometer measurements. Then, bandpass 
calibration, flux calibration, and antenna gain calibrations were carried out. Polarimetric calibration was then done. 
First, the polarization properties of J0510+1800 were roughly estimated by solving for gain ratio of the linearly polarized feeds, X and Y. Then, the cross-hand delays 
and residual $X-Y$ phase were solved for, and the Stokes Q/U ambiguity in the calibration was resolved through examination of the X/Y 
gain ratios. With this, the source properties were determined. The final step was to calibrate the polarization leakage, or $D$-terms. The parallactic angle 
coverage for J0510+1800 was sufficient to allow the calibration of both source properties and $D$-terms. 

For all but one of the targets, Per-emb-41,
we preformed one iteration of a phase-only self-calibration over the integration time to improve the S/N in the images. Per-emb-41, however, was
bright enough to self-calibrate over a 6 second interval. 
We used the CASA task \texttt{CLEAN}, with natural weighting, to produce the final, full Stokes images for all sources. The
typical resolution for our observations was 0.38$\arcsec$ by 0.31$\arcsec$.

Our calibrated images were used to make polarization intensity, polarization angle, and polarization fraction maps. To do this we used
the CASA task \texttt{immath}. The linear polarization intensity map represents the quantity $\sqrt{Q^2+U^2}$, and the polarization angle map represents the quantity $0.5 \arctan(U/Q)$. The polarization intensity maps were debiased using the average noise value determined from the Q and U maps. Each map was masked below 5$\sigma$. The linear polarization fraction map was then formed by dividing the linear polarization map by the Stokes I map.

\section{Results} \label{sec:polres}

\begin{figure*}[htp]
\centering
\includegraphics[width=\textwidth]{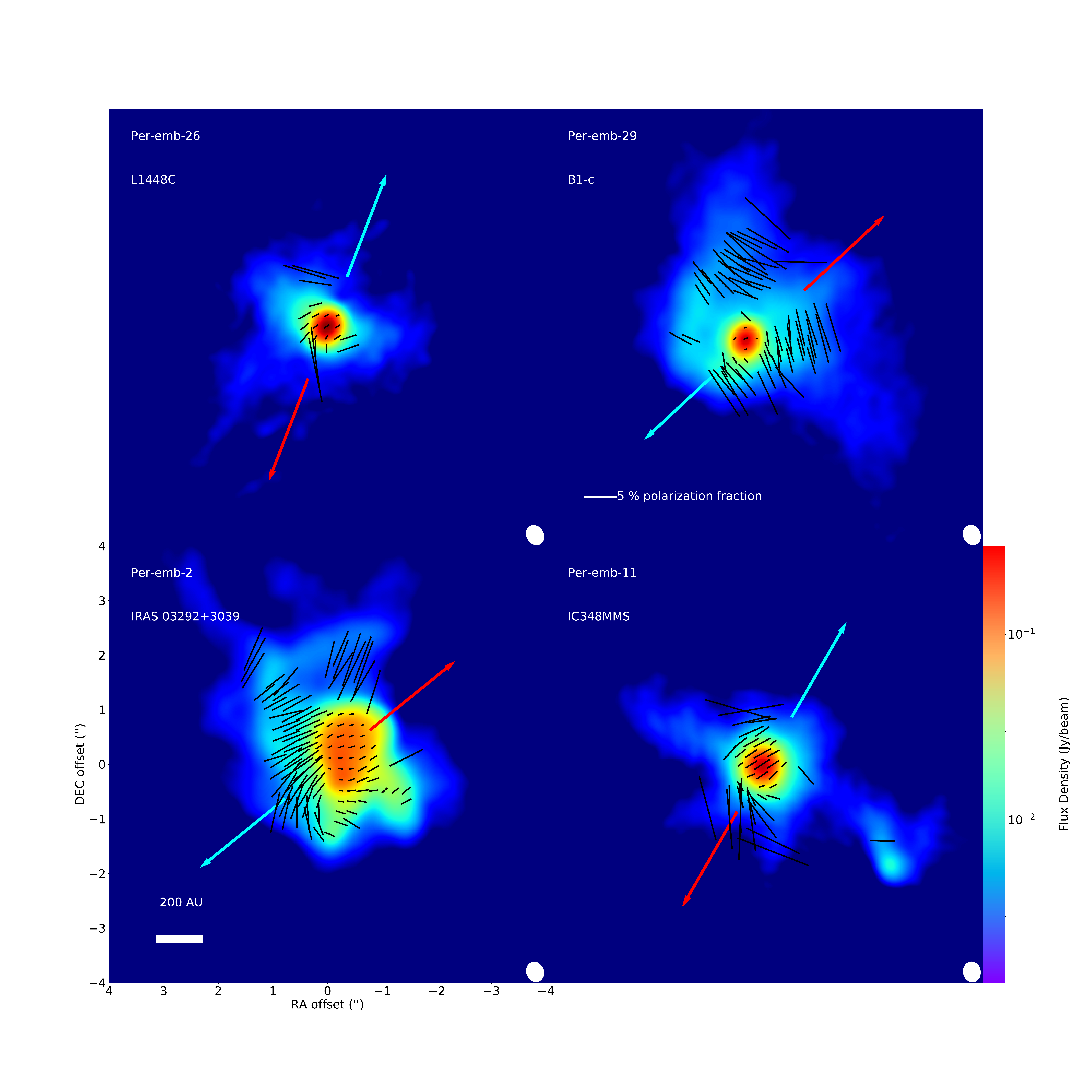}
\caption{870 $\mu$m dust continuum emission is shown in the color-scale.
Black vectors show the (non-rotated) polarization. Their length
corresponds to the polarization fraction. Outflow directions are shown in red and blue arrows, and the beam 
is in the bottom right corner. These four sources have significant large scale emission from their envelopes. These maps were made using emission above 5$\sigma$ ($\sim$ 47 mJy) in polarized intensity.}
\label{fig:pol}
\end{figure*}

In Figure \ref{fig:pol} we show images of our sources. In Table \ref{tab:source} we summarize the 
results. Integrated and peak flux densities are estimated from elliptical Gaussian 
fits using the CASA task \texttt{imfit}. We also present the polarization properties and outflow 
angles (taken from \citealt{st17}, and references therein). We detect polarization at a $\gtrsim 5\sigma$ level in each source.
Because the polarization is relatively well ordered inside of $\sim$ 150 au of each protostar, we present 
the average polarization angle within this region. The average polarization angle is estimated by forming the polarized-intensity weighted average, i.e.
\begin{equation}
<\theta> = \frac{\sum_{i} P_i \theta_i}{\sum_{i}P_i}\\
\end{equation}
where $P_i$ is value of the linearly polarized intensity in pixel $i$ and $\theta_i$ is the polarization position angle in the same pixel. The sum is taken within a 
circle of radius 115 au (= 0.5 arcsecond) centered at the peak of the polarized continuum. 

In all sources, regardless of their classification as disk-candidate/non-disk-candidate sources (via their 8 mm emission), we find a significant difference in the 
small-scale polarization structure near the central protostar and the larger scale polarization structure in the envelope. The morphology and 
percentage of the inner (disk) and of the outer (envelope) regions starkly contrast. To quantify the effects of the differing polarization fraction limits in the inner regions versus the outer envelopes in our maps, 
we computed the lowest detectable polarization fraction in the outer regions. We find that the higher polarization fractions in the outer envelope are not merely marginally detected due to the lower sensitivity to polarization fraction in the envelope. On the contrary, the typical detection is at a level $\sim$ 2 times the minimum detection threshold.

To further quantify this, we used the \texttt{lifelines} Python package \citep{dp17}\footnote{This package is available at \url{https://github.com/CamDavidsonPilon/lifelines/}}  to compute the Kaplan-Meier product limit estimator for the probability 
distribution for polarization percentage. This quantity is essentially a cumulative probability distribution function (CDF) that can 
incorporate upper limits. To construct these, we sample the polarization fraction (or its upper limit) at 
spacings of $\sim 1/2$ beam-width (to ensure a degree of statistical independence). These samples are put into 
two categories: those samples within $\sim$ 100-150 au in projected separation from the protostar and those outside this 
range. The data from all 10 protostars are combined. These distributions are shown in Figure \ref{fig:cdf}.
They clearly show that the inner/outer regions have similar 
distributions of polarization fraction up to $\sim 1-2\%$, but that the outer regions have a significantly higher chance of 
exhibiting a large value ($\gtrsim$ few \%) as compared to the inner regions. The p-value for the log-rank test between 
the two classes is $p \lesssim 10^{-4}$. 

\begin{deluxetable*}{llllrrrrr}
\tablecolumns{9}
\tablewidth{0pt}
\tabletypesize{\scriptsize}
\tablecaption{Perseus Polarization\label{tab:source}}
\tablehead{
\colhead{Object} & \colhead{Alternate Name} &\colhead{RA} & \colhead{Dec}& \colhead{I Peak Flux} & \colhead{Integrated Flux} &\colhead{Polarization} & \colhead{Polarization} & \colhead{Angle offset}\\
\colhead{} & \colhead{} & \colhead{} & \colhead{} & \colhead{} & \colhead{} & \colhead{percentage} & \colhead{angle} & \colhead{from outflow} \\
\colhead{} & \colhead{} & \colhead{(J2000)} & \colhead{(J2000)}& \colhead{(mJy beam$^{-1}$)} & \colhead{(mJy)} & \colhead{\%}& \colhead{($^{\circ}$)} & \colhead{($^{\circ}$)} }
\startdata
Per-emb-26 & L1448C, L1448-mm & 03:25:38.875 & +30.44.05.281 & 284.2 $\pm$ 3.3 & 354.2 $\pm$ 6.6 &   1.0 $\pm$ 0.1              & 123.4  & 35.6\\
Per-emb-50$^\dagger$ &                  & 03:29:07.769 & +31.21.57.098 & 166.4 $\pm$ 0.6 & 192.9 $\pm$ 1.2 & 2.3 $\pm$ 0.1      & 96.1   & 0.9 \\
Per-emb-21 &                  & 03:29:10.668 & +31.18.20.156 & 94.3 $\pm$ 1.2   & 105.5 $\pm$ 2.3 &  1.1 $\pm$ 0.1              & 116.2  & 68.2 \\
Per-emb-18 & NGC1333 IRAS7    & 03:29:11.266 & +31.18.31.087 & 127.4 $\pm$ 4.2  & 365 $\pm$ 16    &   1.1 $\pm$ 0.2             & 97.1   & 52.9 \\
Per-emb-14$^\dagger$ & NGC1333 IRAS4C$$   & 03:29:13.549 & +31.13.58.107 & 115.8 $\pm$ 1.3 & 186 $\pm$ 3.1   &  1.6 $\pm$ 0.1   & 105.2  & 10.2\\
Per-emb-5 & IRAS 03282+3035   & 03:31:20.939 & +30.45.30.252 & 291.6 $\pm$ 3.1 & 501.8 $\pm$ 7.9 &  0.8 $\pm$ 0.1               &  0.0   & 55.0 \\
Per-emb-2 & IRAS 03292+3039   & 03:32:17.923 & +30.49.47.824 & 116.8 $\pm$ 5.8 & 1216 $\pm$ 66   & 1.9 $\pm$ 0.3                & 106.5  & 22.5 \\
Per-emb-29 &B1-c             & 03:33:17.878 & +31.09.31.775 & 193.7 $\pm$ 5.4 & 268 $\pm$ 12    &   5.5 $\pm$ 1.2               & 112.9  & 50.4 \\
Per-emb-41& B1-b             & 03:33:20.341 & +31.07.21.322 & 19.42 $\pm$ 0.14 & 19.42 $\pm$ 0.14 & 1.5$^{*}$                   & 125.2  & 87.8 \\
Per-emb-11$^\dagger$ & IC348MMS         & 03:43:57.067 & +32.03.04.762 & 209.2 $\pm$ 4.5 & 397 $\pm$ 12    &    2.0 $\pm$ 0.2   & 122.5  & 27.4 \\
\enddata  
\label{tab:source}
\tablecomments{$^\dagger$ indicates a source identified as a disk-candidate on the basis of fitting the 8 mm visibility profile \citep{se16}. 
$^{*}$ indicates that \texttt{imfit} failed to converge to a solution for the polarized intensity of Per-emb-41 and this value was 
computed using the peak polarized intensity and the peak Stokes I. Polarization percentage
was found by first running \texttt{imfit} on Stokes I and the polarization intensity maps and then dividing them. The uncertainties are quoted 
in percentage and were found using the respective uncertainties of the fits for the source-integrated values of Stokes I and polarized intensity. 
Since the polarized intensity map of Per-emb-41 was unable to be fit, it has no quoted uncertainty.
Polarization angle values were found using a radius of 0.5$\arcsec$ in each source. Uncertainties in the polarization angle over this range are $\lesssim$1$^{\circ}$.
 The values quoted represent the non-rotated, polarization values, consistent with Figure 1.
} 
\end{deluxetable*}

\begin{figure*}[htpb]
\centering
\includegraphics[width=0.90\textwidth]{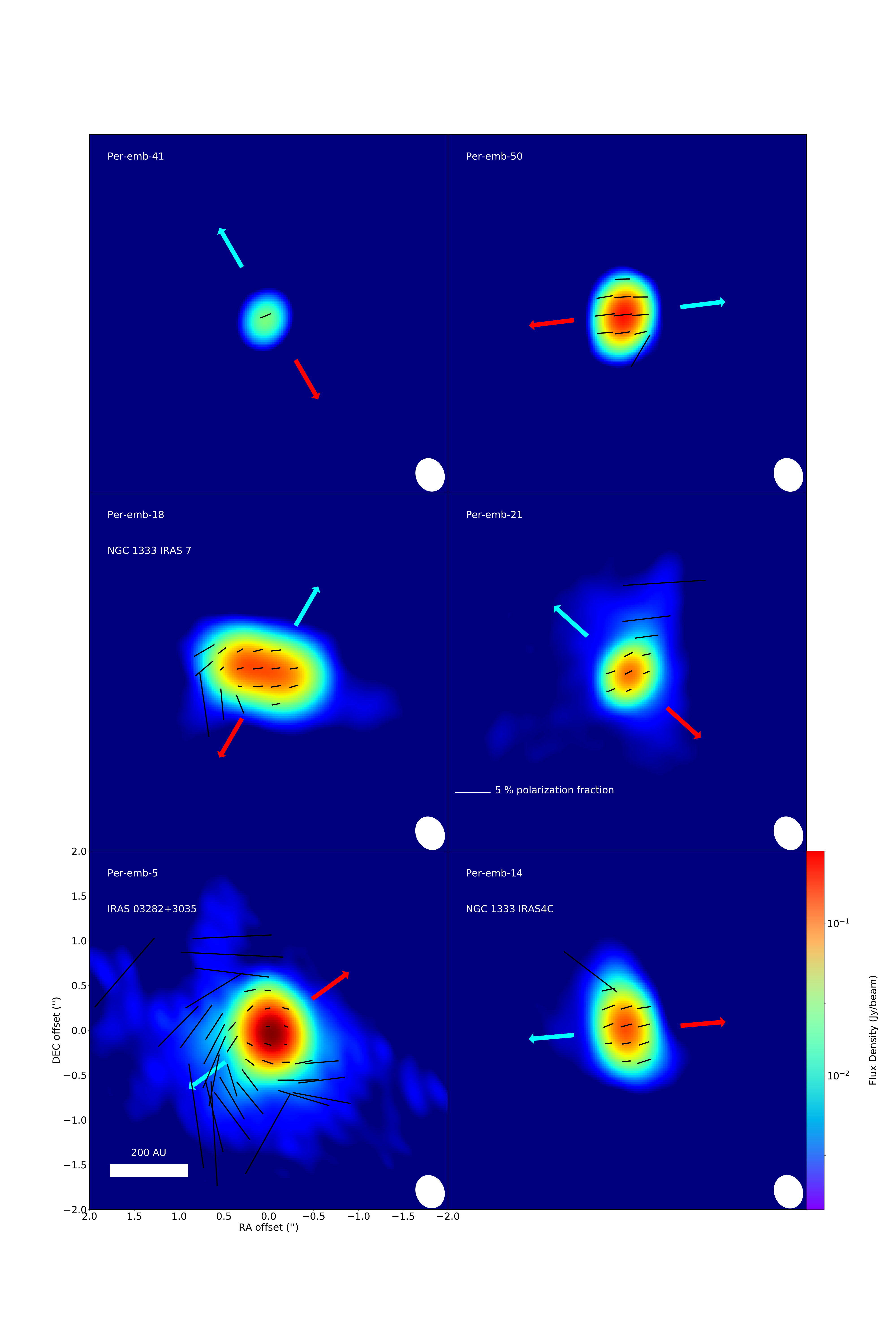}
\vspace{-0.75in} 
\caption{870 $\mu$m dust continuum emission is shown in the color-scale.
Black vectors show the (non-rotated) polarization. Their length
corresponds to the polarization fraction. Outflow directions are shown in red and blue arrows, and the beam 
is in the bottom right corner. These six are more compact and are zoomed in compared to Figure 1. These maps were made using emission above 5$\sigma$ ($\sim$ 47 mJy) in polarized intensity.}
\label{fig:pol2}
\end{figure*}

The angular offset ($\Delta \theta$) distribution of the 
polarization from the 
outflow axis is an observable that relates to the efficacy 
of magnetic braking during the accretion process. 
We performed both the Kolmogorov-Smirnov (KS) and Anderson-Darling (AD) tests to compare the distribution of $ \Delta 
\theta$ to that of a uniform distribution. These results are 
summarized in Figure 4. We find suggestive evidence ($p \sim 
0.02-0.06$) that the angular offset differs from a uniform random 
variable between 0 and 90 degrees for the three disk-candidate sources in our sample and that 
the distribution of angular offsets for the disk-candidate sources is different than for the non-disk-candidate sources. We find no evidence 
($p\gtrsim 0.3$) that the offset distribution for the non-disk-candidate sources is different from a random 
uniform distribution. 

\begin{figure*}[htp]
\centering
\includegraphics[width=\textwidth]{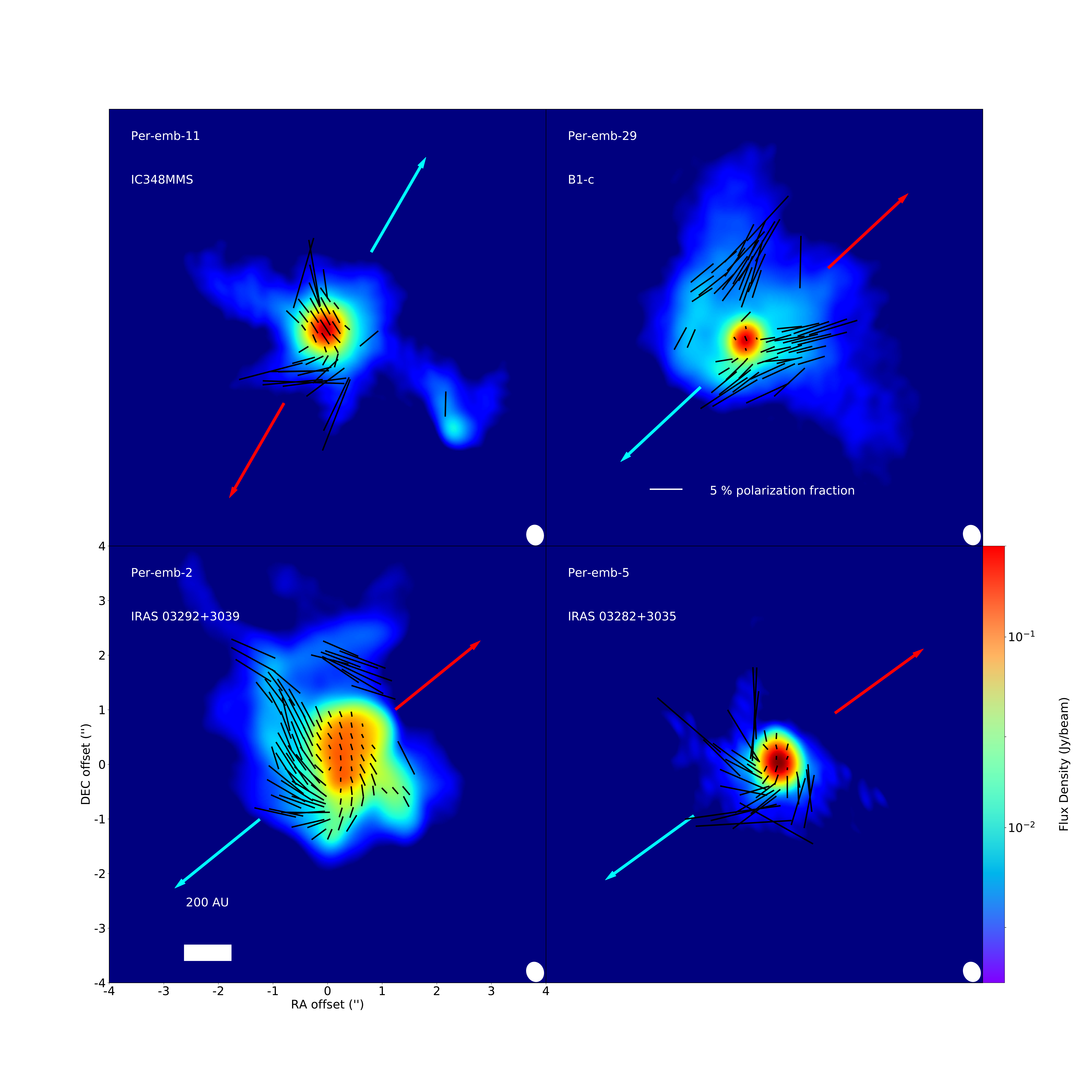}
\vspace{-4.5em}
\caption{870 $\mu$m observations of our sources with significant envelope emission shown in the color-scale. 
In this figure, black vectors are rotated ($90^{\circ}$) polarization vectors, showing the plane-of-sky inferred magnetic field. Their length
correspond to the polarization fraction. Outflow directions are shown in red and blue arrows. The beam is in the bottom right corner.}
\label{fig:complex}
\end{figure*}

\begin{figure*}[htp]
\centering
\begin{tabular}{cc}
\includegraphics[scale=0.55]{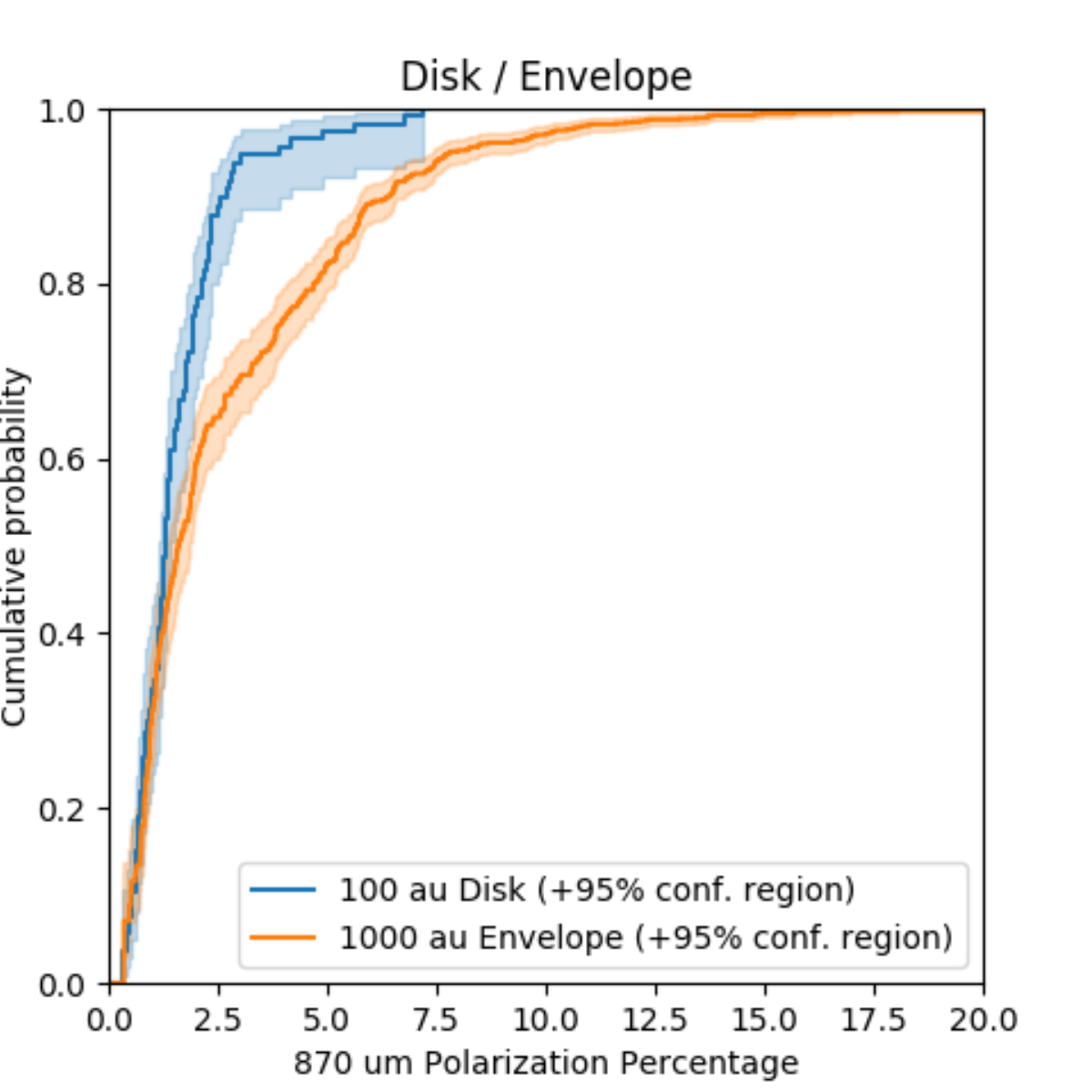}
\includegraphics[scale=0.55]{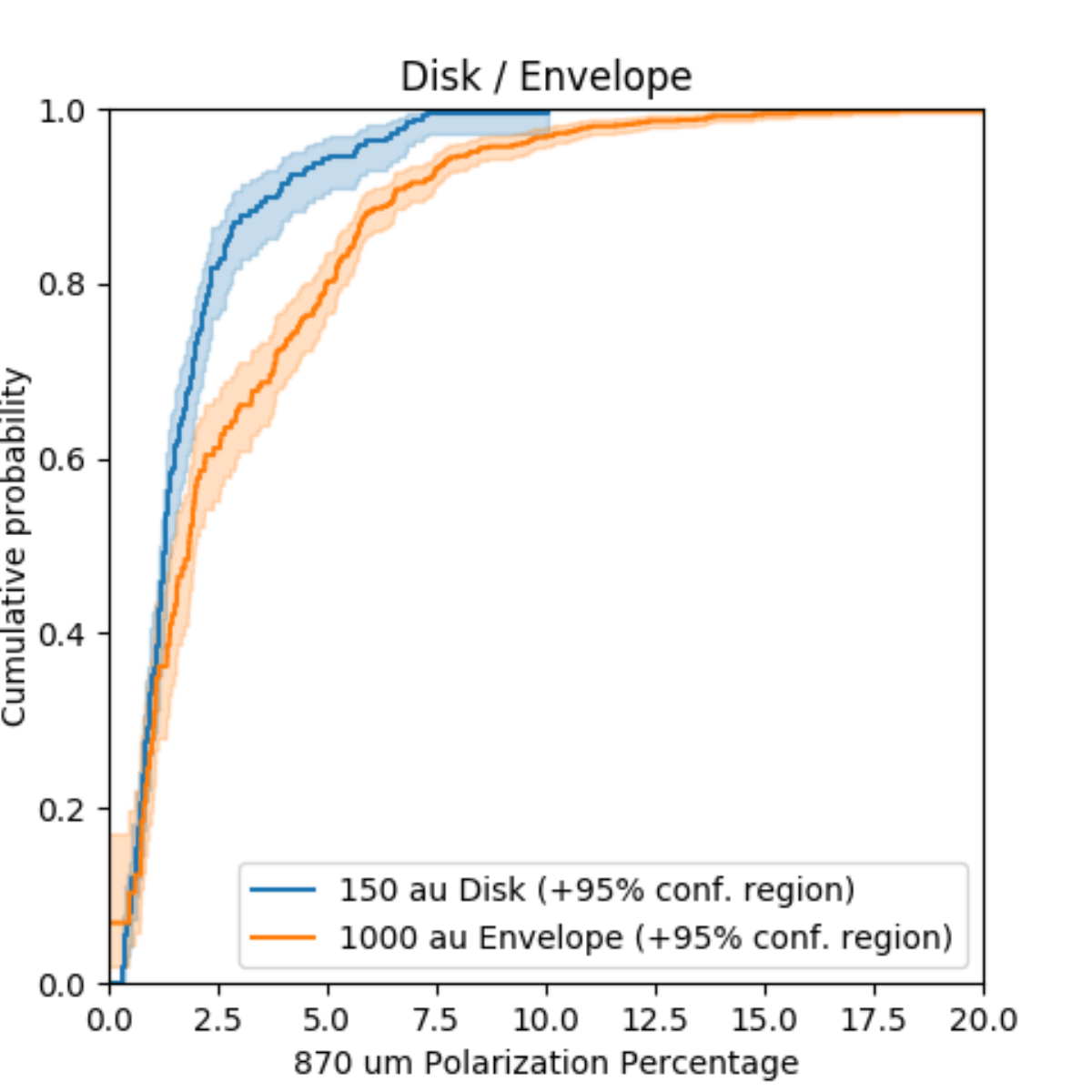}
\end{tabular}
\caption{Disk and envelope KM product limit estimators for two `disk' sizes, 100 au and 150 au, with an `envelope' size of 1000 au. These were made by combining the polarization information for all ten sources.
The shaded area represents the 95\% confidence region for the KM estimator. Note the drastic
difference between the disk and the envelope no matter which radius is used. }
\label{fig:cdf}
\end{figure*}

While the polarization structure of the inner regions (on the disk-scale) of the protostellar 
environment is relatively uniform, that of the envelope is not. In all sources for which significant envelope is detected, we find complex structure in the polarization 
morphology.  Figure \ref{fig:complex} shows the four sources in our sample (Per-emb-2, Per-emb-5, Per-emb-11, and Per-emb-29) with 
relatively complicated envelope polarization. The polarization vectors have been rotated by 90 degrees under the assumption that the mechanism responsible for the polarized emission is magnetically-aligned dust grains. Per-emb-5, Per-emb-11, and Per-emb-29 show 
some morphological similarities to the hourglass morphology expected if the polarization traces magnetic field 
lines dragged in by accreting material \citep[as observed by, e.g.,][]{gi06,ra09,st13}.
In the case of Per-emb-2, the morphology in the center, towards the south, and 
towards the northeast resembles this, but the polarization due east and due north is orthogonal to the expected direction. 

The morphology of the total dust emission from all the sources is also interesting. Very detailed and intricate 
870 $\mu$m dust emission surrounds the young sources. We see a clear distinction between the younger, Class 0
sources and the older, Class I sources (Per-emb-41 and Per-emb-50). Younger sources show a dust envelope which
clearly has structure (see Figures 1 and 2), much of which is 
filamentary or at least not-spherical in nature \citep[e.g.,][]{lo07,to10,le12}. The older sources, on the other hand, 
do not show much envelope emission surrounding the protostar.

\section{Discussion} \label{sec:poldis}
Our principle survey goal was to examine the polarization structure of a set of young protostars to see whether we 
could explain it in the context of models that predict polarization due to different mechanisms, i.e., direct emission from 
magnetically-aligned dust \citep{la07}, thermal self-scattering \citep{ka16}, and direct emission from dust aligned by radiative flux \citep{ta17}.  Our data provide a unique view into this important topic.

The inner $\sim$ 150 au of each source shows a relatively well ordered polarization structure. In some sources 
(Per-emb-2, 5, 26), there is morphological evidence for a smooth gradient in the position angle across the source. In 
others, the polarization position angle is constant to within a few degrees. Especially interesting are the sources with 
candidate 8 mm disks (Per-emb-11, 14, 50), where the constant polarization position angle is aligned (for Per-emb 14 and 50 to within 
about 10$^{\circ}$, and for Per-emb-11 to within 20$^{\circ}$) with both the minor axis of the disk and with  the outflow axis. 
Figure 5 provides quantitative evidence for this. Such a signature is broadly consistent with an 
origin in self-scattering in the disks, wherein the polarization is parallel to the disk's minor axis. 
This morphology may also be consistent with magnetically aligned grains as 
well. While the polarization angle shows virtually no change in the inner region,
these disks' inclinations range from $\sim 45-65^\circ$ \citep{se16}, making it unclear whether the change in polarization angle due to 
grains aligned with respect to a toroidal magnetic field would be obvious. However, the uncertainty in the inclination angle is $\sim 10^{\circ}$, which 
means that it is plausible that these sources are nearly edge-on, in which case no change in the polarization position angle across the source would be observed. It is unclear the degree to which the inner envelope in these sources (as opposed to the disk) is contributing to the polarization signal. In the non-disk-candidate
sources, the polarization is essentially randomly oriented with respect to the outflow axis,
which is very different from the disk-candidate sources. 

\begin{figure*}[htp]
\centering
\includegraphics[width=\textwidth]{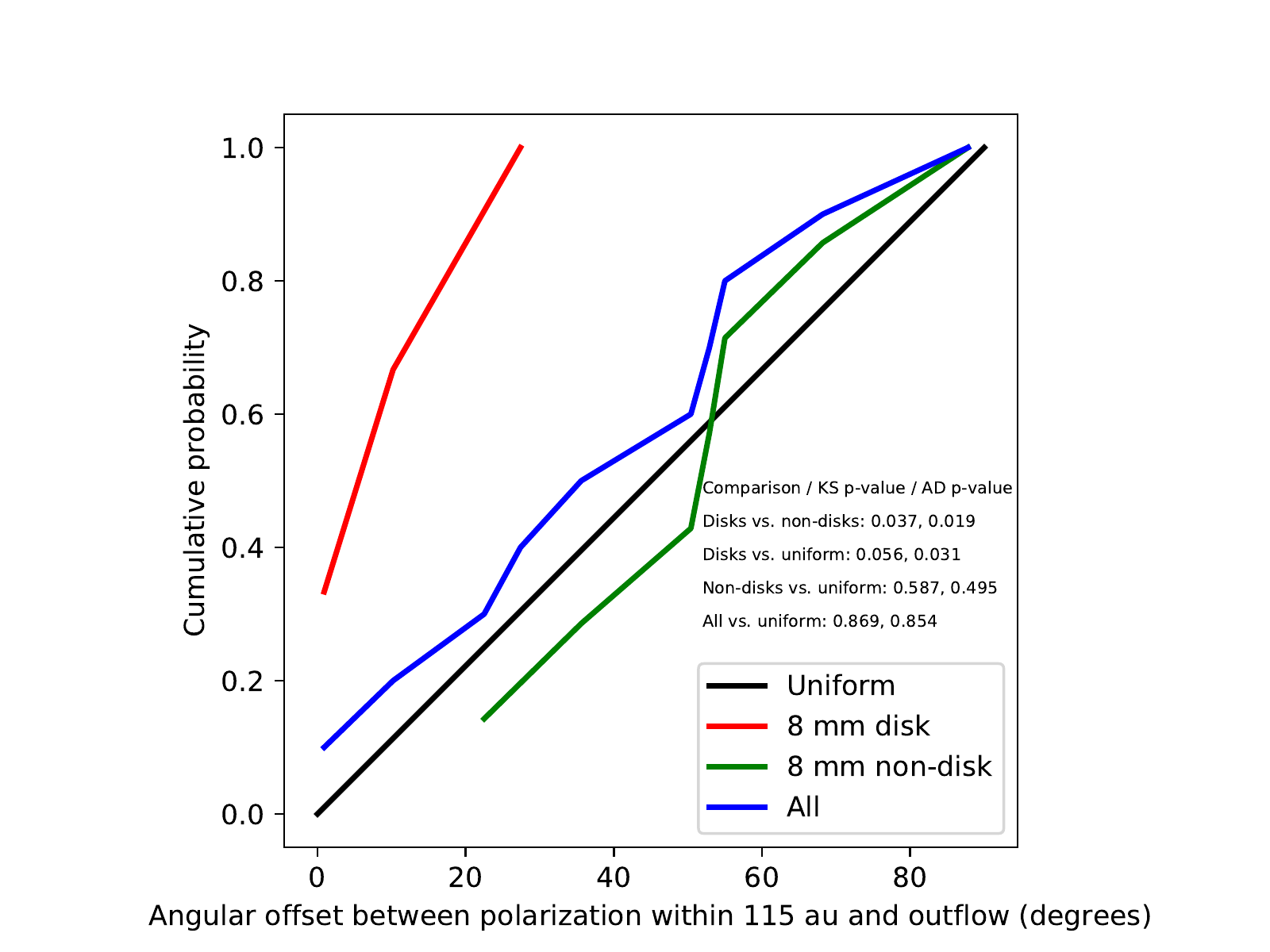}
\caption{Distribution of the angular offset of the small-scale polarization angle and the outflow angle as projected onto the sky. 
Sources are both divided into disk-candidates at 8 mm (red) and those that are not (green), as well as combined 
(blue). 
The distribution for a uniform distribution is shown in black. P-values for the comparison of the disk-candidates/non-disk-candidate/uniform 
distributions for both the KS (left) and AD (right) statistical tests are given. Note that while the disk-candidate line is
quite far from the uniform line, the p-values are high. This is likely due to our small (3) sample size.}
\label{fig:mag}
\end{figure*}

Our data show a stark contrast in the polarization levels  near the protostar and in the
envelope. This `polarization hole' has been observed in both prestellar cores and protostars \citep{do96,ma02,gi06,li13,hu14}. 
This suggests that the polarization in the envelope inherently differs from the polarization
in the region near the protostar. Though we cannot definitively state the cause of this hole, there are a few possibilities. 
One is that the grains are magnetically aligned and that the magnetic field becomes more tangled in the inner regions. 
Such a scenario has been suggested for such polarization holes observed in single-dish \citep{do96} and 
interferometric \citep{hu14} surveys. For our sources, this would imply that the field near the protostar is 
disordered even at $\lesssim$ 20 au size-scales. Another possibility is the difference in grain sizes between the outer envelope and inner envelope/disk. 
It is likely that the grains near the protostar have grown much larger than those in the envelope. It is also 
plausible that small grains are more non-spherical and aligned easily by magnetic fields, 
so it could be that this alignment dominates the envelope polarization. Third, regardless of the mechanism 
that produces the polarization, optical depth effects may play a role. For both scattering and emission from magnetically
aligned grains, the polarization at low/intermediate $\tau$ is higher than that at $\tau \gtrsim 1$ \citep{ya17}, although polarized thermal emission is expected to decline more rapidly than polarization from scattering as optical depth increases. 
Finally, the inner-region polarization could be from a different mechanism, such 
as scattering or radiatively-aligned grains, that could conceivably result in a lower polarization percentage. Self-scattering
is expected to yield $\sim$ 1\% of polarization fraction \citep[e.g.,][]{st14,ya16a}, which is close to what we see in the inner regions.
Alternatively, radiative alignment will have similar fractional polarization to magnetically aligned grains, except close to the central regions, where
the polarization may decrease as the optical depth increases \citep[see Eqn. 5 in][]{an15}.
An interesting possibility is that the areas of low ($\sim 1\%$)
polarization might not be a true rotationally-supported disk, but instead an infalling envelope. 
In this scenario, perhaps both the inner and outer regions harbor aligned grains, just aligned to different extents. If this is the case, 
the rapid change of polarization direction between the inner and outer envelopes observed in our non-disk-candidate sample points to a quite complex magnetic field morphology in the transition region on $10^{2}-10^{3}$ au scales.

\section{Summary}
We have presented our 870 $\mu$m dust polarization survey of 10 protostars using ALMA. These sources consist of 
3 disk-candidates and 7 non-disk-candidate sources. All sources show significant levels of polarized emission, and most show a stark contrast in both
their morphology and polarization percentage between the inner and outer regions. We find evidence that our disk-candidates show 
a polarization signature akin to either self-scattering in their inner region or grains aligned with a toroidal field in an inclined disk, 
while the non-disk-candidate sources show a randomly aligned polarization angle.
We also have shown that since the morphologies and percentage levels in the extended envelope emission 
are very different, and that it may be dominated by another mechanism, most likely magnetically aligned grains.
Additional modeling and multi-wavelength observations are needed to further disentangle the different polarization mechanisms in these young sources.

\acknowledgments
We thank the anonymous referee for their helpful comments.
We thank Katherine Lee and the VANDAM team for invaluable work on the source list. 
This paper makes use of the following ALMA data: ADS/JAO.ALMA\#2015.1.01503.S. ALMA is a 
partnership of ESO (representing its member states), NSF (USA) and NINS (Japan), together with 
NRC (Canada), NSC and ASIAA (Taiwan), and KASI (Republic of Korea), in cooperation with the 
Republic of Chile. The Joint ALMA Observatory is operated by ESO, AUI/NRAO and NAOJ. The NRAO is a facility of the National Science Foundation
operated under cooperative agreement by Associated Universities, Inc.
\software{CASA (v4.7.0; \citealt{mc07}), lifelines \citep{dp17}, Matplotlib (http://dx.doi.org/10.1109/MCSE.2007.55)}

\end{document}